\documentclass[12pt,A4,oneside]{article}
\usepackage{graphicx}% Include figure files
\usepackage{dcolumn}% Align table columns on decimal point
\usepackage{amsmath,amsfonts}% popular packages from the American Mathematical Society
\usepackage{bm}% Bold Math package
\usepackage{setspace}
\usepackage{subcaption}
\usepackage{natbib}
\usepackage[hmargin=1.25cm, top=2cm, bottom=2cm]{geometry}
\usepackage{hyperref}
\hypersetup{
    colorlinks=true, %set true if you want colored links
    linktoc=all,     %set to all if you want both sections and subsections linked
    linkcolor=blue,  %choose some color if you want links to stand out
		citecolor=blue
}
\usepackage{caption}
\usepackage{fancyhdr}
\graphicspath{ {images/} }
\DeclareGraphicsExtensions{.jpg,.png,.pdf,.eps}
\makeatletter
\renewcommand\paragraph{\@startsection{paragraph}{4}{\z@}%
            {-2.5ex\@plus -1ex \@minus -.25ex}%
            {1.25ex \@plus .25ex}%
            {\normalfont\normalsize\bfseries}}
\makeatother
\usepackage[utf8]{inputenc}
 
\begin{document}

\title{Auditory information loss in real-world listening environments}% Force line breaks with \\

\author{Adam Weisser\\Macquarie University, Sydney, Australia}

\date{\today}% It is always \today, today,
\maketitle 

\abstract
Whether animal or speech communication, environmental sounds, or music -- all sounds carry some information. Sound sources are embedded in acoustic environments that contain any number of additional sources that emit sounds that reach the listener's ears concurrently. It is up to the listener to decode the acoustic informational mix, determine which sources are of interest, decide whether extra resources should be allocated to extracting more information from them, or act upon them. While decision making is a high-level process that is accomplished by the listener's cognition, selection and elimination of acoustic information is manifest along the entire auditory system, from periphery to cortex. This review examines latent informational paradigms in hearing research and demonstrates how several hearing mechanisms conspire to gradually eliminate information from the auditory sensory channel. It is motivated through the computational need of the brain to decomplexify unpredictable real-world signals in real time. Decomplexification through information loss is suggested to constitute a unifying principle of the mammalian hearing system, which is specifically demonstrated in human hearing. This perspective can be readily generalised to other sensory modalities.

\section{Background} 

Various types of acoustic information are regularly invoked in hearing research -- e.g., spectral, temporal, spatial, envelope, intensity, speech -- yet information theory itself \citep{Shannon} is generally not introduced as a milestone of the field. Nevertheless, the explicit influence it has had on the related fields of cognitive psychology and neuroscience, and the more implicit roles in psychoacoustics and signal processing, cannot be overstated. %The same can be said about the inclusion of the more gradually evolved computation theory, which deals with the processing of information. 
The split between formally introducing and implicitly harnessing information in the various auditory domains may be understood given the conceptually strict way in which information is defined and used, and the historical backlash against applying it loosely \citep{ShannonBandwagon}. Additionally, projecting information theoretical concepts on the operation of the human brain is not universally accepted, even if it dominates neuroscience \citep{Searle}. Still, it is possible to reinterpret a very broad range of auditory phenomena as motivated by information processing economy, without entering any of these formal debates, by considering hearing as a communication system operating in real-world conditions. 

% Read: Buracas, Adelman, Eggermont, Theunissen
A brief review of some concepts from information theory that are pertinent to hearing is presented in Section \ref{IT}, followed by a breakdown of auditory research paradigms compared to real-world acoustic environments in informational terms in Sections \ref{paradigms} and \ref{realworld}. The role of the auditory channel is then contextualised within the general information processing of the brain in Section \ref{brain}. It is then argued in Section \ref{mechanisms} that auditory information loss is a necessary goal and not a side-effect of the system. Lossy compression of speech is given as a concrete example in Section \ref{speech}. The lossy perceptual framework is suggested to be applicable in other modalities in Section \ref{OtherMods}.

\section{General concepts of information}
\label{IT}
Below is a qualitative introduction to information theory, relevant to hearing research. Comprehensive treatments of the subjects are found, for example, in \citep{Shannon,Pierce,Cover}. 

A generic communication system comprises a source that transmits a message to a receiver, through a physical channel \citep{Shannon}. The amount of information that a given message carries corresponds to the level of uncertainty or unpredictability that it has, relative to the ensemble of all possible messages. Thus, the degree of uncertainty (quantified using Shannon’s entropy and measured in bits) is computed solely based on the probability distribution of the messaging system, which is independent of the meaning of the messages. The messaging convention, or its code, is unique for the transmitter-receiver pair and the context of their communication. 

Messages are transmitted over time in sequences of symbols over the information channel. The channel is generally susceptible to noise that can lead to ambiguous reception and results in errors. However, it is possible to combine individual symbols in sequences (codewords) so that the reception error caused by the noise is made arbitrarily small, at the cost of a lower rate of information transmission on the channel. This is achieved by increasing the amount of redundancy in the code -- repeated information that can be used by the receiver to disambiguate the reception in noisy channels. Alternatively, the code can be made more efficient by removing redundancies in a process called lossless compression, which can significantly decrease the size of the message, but make it more susceptible to decoding errors and ambiguity. A more aggressive process is called lossy compression, whereby non-redundant information is removed and cannot be recovered, even when the message retains its intelligibility. Because of its physical nature, the channel has a finite capacity to carry information per unit time, which influences what code should be used, how efficient it should be, and what kind of compression is entailed by it. As information can flow between systems and channels, codes and physical constraints change throughout, but the amount of information in a channel can never exceed the amount in the preceding channel.

The concept of redundancy can be taken a step further than in Shannon's information theory, where patterns are compressed based on their relative probabilities in the ensemble. Algorithmic complexity (also referred to as Kolmogorov complexity, \citealp{Kolmogorov}) is defined as the shortest computer program that can be written on a universal Turing machine (a computer programmed with a generic language; \citealp{Turing}), which reproduces the pattern. Algorithmic complexity asymptotically matches the entropy of the sequence when it cannot be compressed further (i.e., when it is random). Importantly, algorithmic complexity is not universally computable, i.e., there is no general way to find the shortest computer program that can reproduce an arbitrary sequence, except for trying all programs possible (see discussion in \citealp[p. 482-484]{Cover}).

While messages are conveniently treated as made of discrete symbols and sequences, all communication has to be realised in continuous physical means. Physical systems, however, have unbounded bandwidth and dynamic range, which means they produce infinite amount of information. Fortunately, it is always possible to transform between the continuous and the discrete representations by bandlimiting the communicated signal, and quantizing its instantaneous levels to fixed steps through sampling. This discretisation may be done at any level of precision (or reproduction error) to represent the original signal using the proper choice of bandlimiting and quantization. Without loss of generality, the bandlimited continuous signal can then be realised by modulating a fast carrier wave with slow changing information in its complex amplitude \citep[e.g.,][237-241]{Couch}. 

\section{Hearing research paradigms in informational terms}
\label{paradigms}
The concepts of information theory -- transmitter, receiver, and channel -- have not been formally defined in hearing research, except for sporadic applications. Yet, the informational framework is necessary in order to account for information flow and loss produced by auditory mechanisms, contrast them with real-world conditions, and relate them to brain theories. Therefore, several standard experimental concepts, methods, and tools in hearing research are functionally analysed below in informational terms. This analysis does not pertain to any specific experiment, but is abstracted from standard practices of the field \citep[e.g.,][]{Moore2013, Gelfand}. %While hearing is especially well-suited to deal with communication signals, it does not depend on the source intentionality to be received, since passive environmental sounds are equally well-received. Therefore, `communication' must be taken in the most general way, as a process of information transfer between a sound emitting object and a listener. 

\subsection{The sound source}
The message-containing source, target signal, or stimulus, is typically produced electroacoustically using loudspeakers, or headphones, and seldom using live sources (e.g., talkers, musical instruments). Unlike played-back recorded material of real sources, artificially generated stimuli offer superior experimental control. The building blocks of hearing-test stimuli have traditionally been the pure tone, noise burst, and impulse, and many variants thereof obtained by filtering, modulation, and superposition. These sounds carry only the information required to elucidate specific aspects of the system operation. For example, a (realistic, finite) pure tone is generated using five pieces of information: frequency, amplitude, onset and offset times, and initial phase. Therefore, it very low algorithmic complexity, given an algorithm that can generate sinusoids. Such synthetic sounds are considerably less informationally rich and complex than naturally produced ecologically-relevant vocalisations \citep{Theunissen}. In any case, the choice of source constrains the kind of information that the listener must auditorily process.  %Here all signals are computed discretely, performing equivalent functions to signals obtained by analogue signal theory, mainly using Fourier analysis for periodic sounds, and correlation analysis for stochastic signals. Generating these signals is made precise using modern digital-to-analogue converters, which feed continuous electrical signals to the audio systems that produce the signals that get to the ears. 

\subsection{The sound receiver}
The listener's auditory system, or any of its subsystems, is the recipient of the messages from the acoustic source. Its role may be completely passive, as when acoustic properties of the ear are measured (e.g., ear canal transfer function, otoacoustic emissions). It can also demand active listening when decision-making about the sound is required (e.g., threshold measurement, speech intelligibility). The signals are often designed in an attempt to tap only part of the auditory system, so that the information they carry targets only that part and interacts with other parts in a predictable manner. For example, if a stimulus is identical but is out-of-phase on the two ears, then whatever effect this interaural phase difference has must be the result of binaural processing. Since the auditory system in all mammals is very similar -- albeit with different tuning -- most in-vivo physiological data (and some psychoacoustic data as well) about it have been obtained using animal models, and the knowledge is assumed valid for humans as well in many cases \citep{Long}. The source-receiver pair forms the basic communication system in hearing that is mediated via the channel.

\subsection{The information channel}
The information channel depends on the input and output points of the measurement, and it is usually a cascade of several distinct media. In the case of loudspeaker presentation, the room is the immediate medium. Acoustic reflections and reverberation from the room boundaries distort the signal and are preferably removed by situating the system in an anechoic chamber or a soundproof booth, but may also be desirable in some cases. Either way, the experimenter has a choice in demarcating the stimulus and channel. %Air is negligibly absorptive in the audible bandwidth (but is more significant for ultrasonic communication), while water has more complex propagation properties \citep{Kinsler}. 
In headphone presentation, sound travels almost straight into the auditory periphery, where in a series of transductions, information travels up the auditory pathways. Acoustic inputs are then bandpass-filtered to narrowband channels, or auditory filters, which retain their order tonotopically along the auditory pathways \citep[e.g.,][]{Merzenich1974}. Typically, only information involving neuron firing is explicitly referred to as `coding', where the temporal, level, speech, and other auditory codes are still being studied \citep{Eggermont,Sayles}. 

Regardless of the auditory channel segment that is involved in the test, all internal and external noise and distortion sources in the system should be accounted for, to eliminate measurement biases of the receiver's response. Other communication features should be controlled, as the availability of information running on a parallel channel (e.g., visual \citealp{Sumby}), or the existence of feedback or memory, complicates one-directional information flow assumption \citep{Marko}.

\subsection{The task}
The experimental task ties together the stimulus, channel, and listener, to bring out only certain aspects of hearing. In active tasks, decision-making is required, which implies that an underlying mechanism must process the input and extract certain information from it. This is true regardless of the operation -- detection, recognition, comparison, identification, estimation, etc. -- all require some computation. It may include hard-coded operations such as filtering, or neural recoding, but can also be much more elaborate, such as recognizing speech in noise and reverberation. Finally, behavioural studies are administered in trials separated by pauses, which allow listeners to momentarily rest and sometimes respond not in real-time.

\subsection{Functional roles}
The functional roles that controlled stimuli and tasks fill for the receiver are determined by design. Ideally, the message from the information source and its role for the receiver are completely known by the `omniscient' experimenter \citep{Eggermont}. Stimuli are designated by the experimenter: the roles of target, noise, masker, distractor, figure, background, etc. In and of themselves, the stimuli are almost always meaningless for the subject, whose response space is confined to one or few task-relevant dimensions with well-defined range. Usual tasks are designed to test the capability of the auditory system to extract relevant information despite poor signal-to-noise ratios, competing signals, or all too subtle cues. The performance in the task can also be presented as success or failure amid challenging conditions, where in some cases, confusion, distraction, fatigue, mind-wandering, or other high level extraneous factors affect the listener. 

\section{Real-world environments and listening}
\label{realworld}
Everyday listening entails fundamentally different situations to those set up in laboratory-based experiments \citep{Theunissen}, as are prescribed by the experimental paradigms above \citep{Buracas}, and may entail situations that could potentially overwhelm the listener in comparison. Imagine being in a small and busy street: every minute dozens of cars of different size move in different directions and speeds. People are walking on sidewalks, coming in and out of stores, chatting, shouting, or eating in public. Occasional sounds of music and machinery come from within the buildings. Some passers-by carry devices that emit all sorts of electronic sounds. Birds are calling from the treetops. Occasionally, an aeroplane flies above, or a siren goes off. All sounds reverberate between the building facades, and happen more or less concurrently. While each aspect of this acoustical display is completely controlled in a laboratory setting, it is not the case in the actual street.

Several differences to laboratory-based situations are evident. It might be tempting to refer to such a scene simply as noise -- a collection of unwanted sounds \citep[cf.,][p. 273]{Schafer} -- but this sound contains rich information about the street and the life within it. Moreover, the number of acoustic sources in such an environment are generally unknown and unbounded, so  the listener cannot have complete knowledge of the identity of all of them. Also, the listener has no absolute control over what kind of sounds they will generate, which means that auditory inputs are inherently unpredictable and cannot be forced into a closed categorical response space. As a corollary to the continuous nature of real sources (Section \ref{IT}), the natural acoustic environment contains an infinite amount of information. Importantly, listeners have no predefined `task', but rather continuously negotiate what source(s), if any, should count as an instantaneous `target', while some undesired sound(s) as `noise'. This happens automatically (unconsciously) and dynamically, while other external and internal stimuli that are not necessarily acoustic compete for the listener's attention. Therefore, the finite-resourced listener must select what information, if any, to process, whereas unused information is filtered, suppressed, ignored, deselected, or turned into noise. 

\section{Brain theories}
\label{brain}
The auditory system itself constitutes only one channel of perceptual information that feeds into the brain's decision making processes. As the brain negotiates signals from different internal and external inputs, infinite information rates from hearing  may be unsuitable for efficient real-time operation of the finite brain without some preprocessing of the inputs by the auditory system itself. Rather disturbing evidence for the possibility of computationally overloading the system is evident from a review by \citep{Lipowski} of studies that subjected listeners to stimuli of chaotic displays of loud noises and bright lights. The resultant sensory stimulation was so overwhelming for some individuals, that they experienced heightened arousal, anxiety, sadness, aggression, paranoia, and even involuntary sleeping and hallucinations\footnote{These tests undoubtedly placed significant emotional stress on the subjects -- something which could have interacted with the mere perceptual overloading effects.}. As extreme as these conditions are, they are most definitely something to be avoided for healthy survival. 

Fortunately, auditory information is finite. Hearing is bandlimited and has a finite dynamic range -- set between the system internal noise (spontaneous neural firing) and the threshold of pain, where the cochlear hair cells get mechanically damaged. These physical constraints yield an upper bound on the auditory input information rate. Furthermore, neuronal recordings of modulation frequency along the auditory pathways (a proxy for information rate) suggest that the maximal rate gradually decreases on the way to cortex \citep[Figure 9]{Joris}, so information is eliminated or compressed the higher up signals are neurally recoded in the ascending auditory pathways. How is this auditory information reduction is achieved?

%Experiments in hearing as are epitomized above show that the auditory system is inherently adaptive, as it has a complex set of mechanisms that allows it to shift its focus in time, frequency, and space to a scale of interest dictated by cognition and action. Microscopically focusing on selected parts of the auditory input requires more information to be processed locally to satisfy these demands, which may come at the expense of information that can be processed from other portions of the input stimulus. Either way, the auditory system must be able to deal with infinite flows of information from the environment. Additionally, it should instantaneously extract cues from the environment in an adaptive manner, which can inform decision making and action of the specific organism, facilitating its survival. These considerations have implications on theories that deal with information processing in the brain, which are briefly touched upon below and motivate the ensuing discussion about hearing.

According to the influential Efficient Coding Hypothesis \citep{Attneave, Barlow}, sensory information from the environment naturally contains redundancies that are gradually removed as the neuronal signal gets to cortex and eventually to consciousness. The hypothesis was reiterated by \citep{Barlow2001}, stressing that redundancies are critical, but not because they are removed, but rather because they are used to generate statistics -- an internal model about the world that provides the priors necessary for prediction and error correction of future sensory information\footnote{In fact, redundancies are thought to not be compressed in vision \citep{Barlow2001}, but they may be in audition \citep{Chechik, Smith}.}. Redundant patterns are then represented in multiple neurons rather than eliminated by lossless compression (see also \citealp{Chater2003}). 

However, it is not possible to universally apply lossless compression, for arbitrary degree of complexity of realistic stimuli, since no single algorithm exists that can identify hidden redundancies. Thus, it is argued here, there is no reason to assume that the brain is always able to detect those redundancies, if they require more intricate computational processes to uncover than mere statistical pattern detection. In other words, because finding the algorithmic complexity of a sequence is an intractable problem, the brain should be not be capable of reducing just any input signal to its most efficient form and always simplify its representation losslessly. Instead, as will be illustrated in Section \ref{mechanisms}, it uses a lossy compressive toolkit that aggressively reduces the information rate and complexity of stimuli from the environment, while still allowing survival. It explains why the realistic street scene (Section \ref{realworld}) does not acoustically overload the normal functioning individual, unlike the completely chaotic stimuli that are potentially irreducible and incompressible \citep{Lipowski}. This is a demonstration of a rate-distortion theoretical perspective\footnote{Rate-distortion theory is the branch of information theory the is concerned with optimizing the received errors in finite-capacity channels \citep{Shannon,Cover}.}, which showed that there exists a trade-off for organisms between the perceptual precision and its energetic cost due to processing \citep{Dedeo}. According to this view, perceiving the environment in high fidelity is useful only inasmuch as it promotes survival, where additional, costly information can be safely discarded. 

In another relevant perspective, predictive coding theories emphasise how the brain tries to minimise the surprise from perceptual inputs, which can be done using adequate statistical modelling of the world, combined with action \citep{Clark}. While certain variants of this theory are framed in terms of precision-complexity trade-off \citep{Friston}, in many realistic cases it would be incoherent to talk about active predictive modelling of the world based on information that never reaches the predicting system, due to its finite input aperture. Instead, by restricting information along the perceptual pathways the system is heuristically geared either to not bother about missing information, or to accept it as ambient noise. Only when the stimulus results in error signals that cannot be ignored, does the brain have to respond by acquiring better perceptual information, or forming more precise models about these otherwise neglected aspects of the world. 

Considering the brain function along with real-world information content, it is argued here that information reduction is a requirement, rather than a side-effect of perception, in order to avert informational overload on the brain. This is in line with what many auditory mechanisms achieve, as can be appreciated if their operation is reframed as follows: Instead of focusing on the failure to extract information from synthetic stimuli, the successful elimination of information in complex scenes is emphasised, because it relieves the system from avoidable and costly information processing. This is demonstrated in the next section.

\section{Auditory mechanisms}
\label{mechanisms}
Interpreting how auditory processing of real-world signals operates based on known mechanisms from controlled laboratory observations is not trivial for several reasons. First, applying these mechanisms quantitatively requires complete knowledge of all sound sources in the environments, their absolute and relative levels, and their time-dependent relative positions -- knowledge that is generally unavailable and, as was argued above, unknowable. % Moreover, sounds tend to not repeat exactly, or not at all, so they are not reproducible. 
Second, knowledge about auditory mechanisms was obtained using synthetic stimuli that are rarely -- if ever -- found outside of the lab. Inferring instantaneous responses to realistic sounds based on responses to idealised pure tones and noise bursts (for example) is unintuitive and complex. Arguably, this classical approach is of limited value for dynamic signals such as speech (e.g., \citealp{Sharma}). Third, many mechanisms have counter-mechanisms, e.g., masking phenomena have masking release, masked sounds can be glimpsed at or restored, grouping and segregation in scene analysis, attention can switch or remain focused both due to competing processes -- top-bottom processes driven by context, or bottom-up driven by the salience of different events. Observations about when these counter-effects are expected are usually available for particular conditions only (e.g., spatial release from masking for certain targets that are of equal levels and exactly $0^\circ$ or $90^\circ$ apart, but not for arbitrary targets, levels, and locations). 

\begin{figure}
		\centering
		\includegraphics[width=0.6\linewidth]{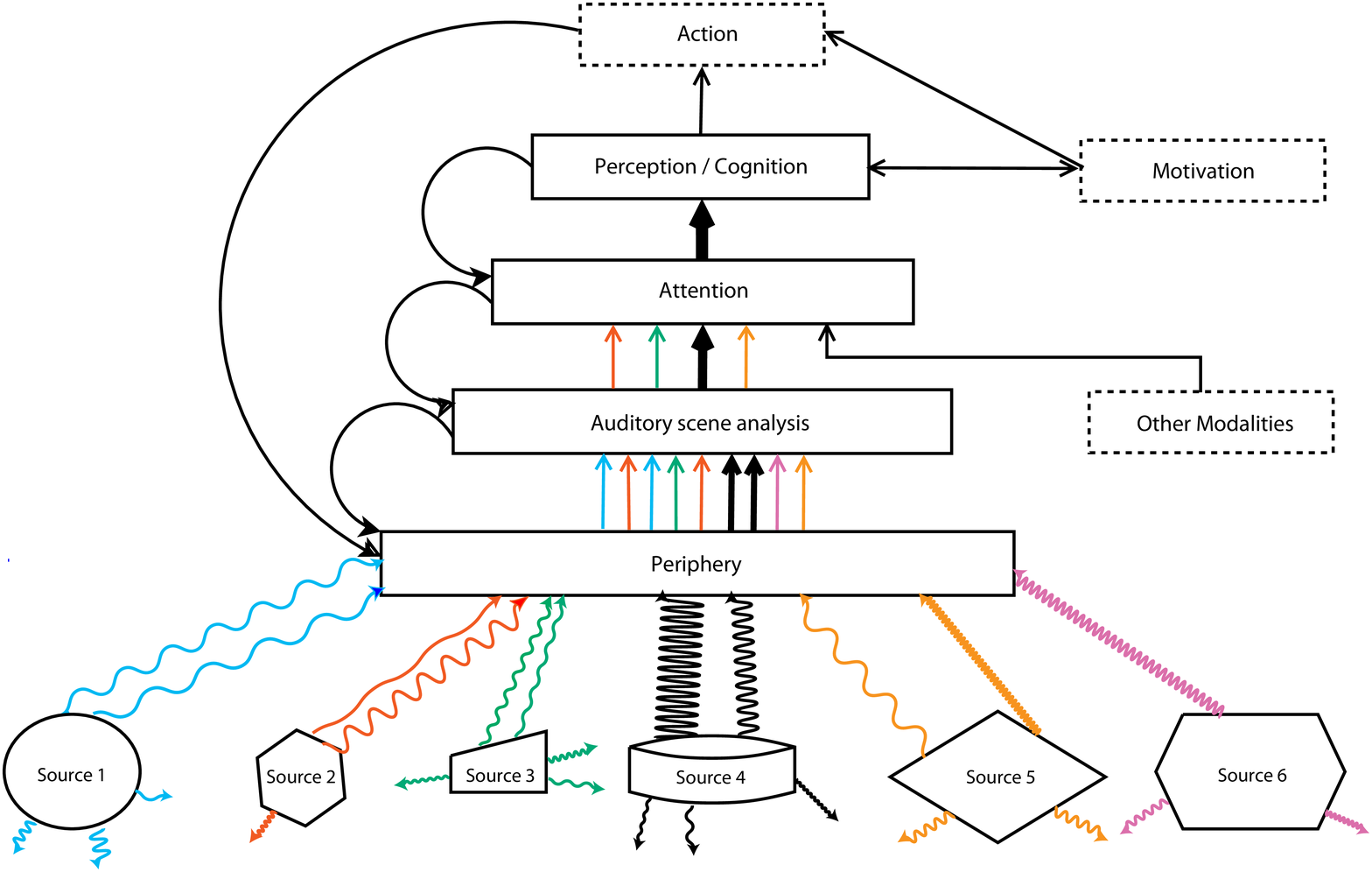}
		\caption{A conceptual model of auditory information loss with multiple acoustic sources. The wavy arrows designate flow of acoustic information of arbitrary nature (e.g., spectral, temporal, spatial) transmitted from the acoustic sources. The number of arrows per source qualitatively indicates the amount of information it transmits -- not all of it is picked up by the ears. Received sounds are coded to auditory information (straight lines) and compressed (less arrows per source / colour) as it travels up the auditory pathways. Source 4 exhibits salience (arbitrarily), so its respective black arrows dominate perception. On a higher level, perception / cognition receives the input from attention and inform action, which may also be modulated by motivation. Action, in turn, can affect what information reaches the periphery (e.g., by head turns). Other general feedback mechanisms are illustrated between perception / cognition and attention (e.g., context) and attention and scene analysis (e.g., switched attention), and back to the periphery.}
		\label{fig:LossModel}
\end{figure}

Figure \ref{fig:LossModel} shows a conceptual model of acoustic information flowing into the auditory periphery, through its ascending pathways, all the way to cognition and perception. Gradually, with every processing level, information is eliminated from the original input (cf. \citealp{Buracas}). Feedback mechanisms exist between all levels that can adapt the system response to the instantaneous needs of the listener, also considering non-auditory inputs. Each processing level encompasses several mechanisms that can be observed using particular experimental paradigms. However, some mechanisms may overlap in function and their physiological realization is distributed over different auditory circuits, in a way that is not always fully understood. A subset of these mechanisms is described below with emphasis on their informational lossiness, but a more comprehensive set is provided in \citep[p. 143-162]{Weisser}. None of the phenomena is exclusive to humans, unless otherwise stated.

\subsection{Low-level mechanisms}
The first group of mechanisms contains phenomena that are largely associated with the periphery (the outer, middle and inner ears, and the auditory nerve), but are not confined to it, as some mechanisms have correlates in neural auditory circuits. Limited bandwidth and dynamic range, which have already been mentioned above, are the two fundamental properties that turn hearing to finite. They are related mainly to the ear mechanics and geometry and are species-specific \citep{Heffner}. 

The two ears are sometimes treated as two independent channels of acoustic information \citep[e.g.,][]{Cherry}. In reality, natural signals tend to be highly correlated between the two ears, and sound events are experienced to be unitary -- auditory information is used to identify and to localise acoustic events in space \citep{Rauschecker2017}. Hence, acoustic sources in space can be efficiently expressed using a monaural signal along with additional interaural parameters (time, phase, or level differences) or functions (head-related transfer function, interaural coherence). Importantly, the binaural parameters are also updated relatively sluggishly \citep{Grantham1978}. This suggests that correlated left- and right-ear time signals can be compressed to a single signal plus a low-rate spatial transformation (e.g., \citealp{Johnston}). %Nevertheless, there are ample contralateral neural projections between the left and right auditory centers all the way from the brainstem to the cortex \citep{Malmierca, Schofield}. 

%In a similar vein, the precedence effect is the result of localizing an acoustic source according to its first-arriving sound, but not to its reflections \citep{Wallach, Haas}. This can be manifest perceptually in different ways \citep{Litovsky, Brown2015}, as reflections are not perceived as separate events, unless they arrive much later than the direct sound. Instead, the direct and reflected sounds are fused to a single event with slightly different properties (depending on the reflection delay), so that redundant information is effectively suppressed, since the main focus of the listener is the source and its relation to space, rather than the space itself.

More complex loss mechanisms involve multiple sources. `Energetic masking' is generally associated with the cochlea and auditory nerve, and may be classified as simultaneous, forward or backward \citep[Chapter 3]{Moore2013}. All types are characterised by elevation of threshold in specific auditory bands as a result of a masking sound. In informational terms, because masking limits the channel-specific dynamic range, its channel capacity is reduced. With rare exceptions \citep{BenShalom}, masking has not been portrayed as a desirable property of the hearing system, because it results in loss of information. Somewhat confusingly, `informational masking' occurs whenever energetic masking cannot explain a change of threshold, as in speech-on-speech tests \citep{Brungart2006}\footnote{Informational masking has not been studied yet in non-human animals \citep{Reichmuth}.}.%Because the scope of this phenomenon is so inclusive \citep{Kidd2008, Kidd2017}, its exact cause and existence have been debated \citep{Watson2005,Shinn2008}. %One particularly important situation in which the effect of informational masking has been observed is the speech-on-speech paradigm, where speech serves as both target and masker, but it is ensured to not cause energetic masking \citep{Brungart2006}. 

Applying masking to real-world situations is particularly confusing, because the loss from masking is not inevitable, as some acoustic cues can `unmask', or release the signal from energetic or informational masking \citep{Culling,Kidd2017,Kidd2008}. For example, depending on the shape of the animal's outer ears and head \citep{Holt2007}, spatial release from masking can happen when the sources are not exactly co-located (unlike how most masking experiments are set up). Realistic sources are almost never co-located, as it is a physical impossibility, unless they all come from a single loudspeaker. Similarly for other release cues, realistic sources are generally uncorrelated, so their modulations and fundamental frequencies rarely match exactly. Still, it is likely that energetic masking is so dominant in some cases that no cues could release it, especially at large sound level differences between target and masker, or for large incoherent sources that occupy a wide spatial angle. Therefore, information is lost from masking when possible unmasking fails. In this case, sound glimpsing and restoration may be sometimes applied to reconstruct partially corrupted messages \citep[Chapter 6]{Miller1950, Cooke2006,Warren2008}.

\subsection{Auditory scene analysis}

After the acoustic information is spectrally analyzed in the cochlea and gets neurally recoded, bits of auditory information are (re-)grouped to produce contingent auditory objects. This preattentive, complex, and lossy process is called auditory scene analysis \citep{Bregman90}. Spectro-temporal patterns that share the same acoustical properties such as being harmonically related, timbrally identical, having common onset time, or coming from the same direction, are often processed as a single auditory object or event, effectively transforming multi-event / channel input to a grouped perceptual stream \citep{Bregman90, Bee}. %Grouping can be hierarchical and take place on multiple processing levels. %Once grouped, the stream may remain fused even while subjected to various transformations. 
Grouping is a powerful method to cognitively handle high information load \citep{Miller1956}, but it is often lossy, as it is accompanied by the removal of fine-grained details of the time series, which are not needed to maintain the sound event identity over time \citep[p. 182-183]{McAdams1993}. The supremacy of grouped or categorised auditory information can be seen in the categorization of acoustically ambiguous phonemes \citep{Ganong}, in timbral constancy despite acoustic manipulations \citep{Charbonneau}, and in the consistent identification of glass breaking sounds that were heavily tampered with acoustically \citep{Warren1984}. 

If the auditory system creates objects and streams and only selects them later (\citealp{Sussman2007}, but see \citealp{Shinn2017}), then it requires the computational resources to be able to process multiple streams in parallel. However, the number of simultaneous streams that can be handled is typically smaller than four (in humans), depending on the precise stimuli and measurement setup \citep{Kawashima,Zhong,Weller}, so complex scenes may transmit more acoustic information than can be tracked in the scene analysis stage.

\subsection{Attention}

Attention is the mental stage in which competing sensory inputs are singled out to be further processed by cognition, which can drive subsequent decision making and action. William James famously asserted that it is impossible to attend to more than one thing at once, unless it implies very simple or highly automated processes \citep[p. 402-458]{James}. This idea metamorphosed to modern attention theories, which combined it implicitly with the concept of channel capacity, by emphasizing the finite resources that attention has to deal with multiple sensory inputs \citep[e.g.,][]{Broadbent,Kahneman}\footnote{Neither Broadbent nor Kahneman cited Shannon's work explicitly, but both mentioned information theory. However, all modern attention theories were inspired by the seminal cocktail party problem \citep{Cherry}, where Shannon's work on written English redundancy was explicitly cited \citep{Shannon1951}. Information channels are still central concepts in the attention literature \citep[e.g.,][]{Pashler, Wickens}.}. Thus, attentional suppression of auditory streams may be the only mechanism that is openly acknowledged in literature to have a desirable role in eliminating auditory information \citep{Golumbic, Ding}. Attention appears to gradually evolve or emerge in the ascending auditory pathways \citep{Shinn2017}, while unattended streams may have to be suppressed to not interfere with the attended signal \citep{Fritz}, as was found in several auditory evoked potential studies in humans \citep[e.g.,][]{Horton, Kong}. Under certain conditions, attention modulates peripheral responses via the medial olivocochlear efferents \citep{Heivet}, and is likely modulated in itself by higher-level factors, such as motivation, as was shown in vision \citep{Balcetis} (see Figure \ref{fig:LossModel}).

In real environments, not all acoustic sources are equal and some are more salient (e.g., louder, having impulsive onset), which can serve as cue for capturing attention through bottom-up processes \citep{Koch1987, Kayser2005}. This comes at the expense of other streams due to a `Winner-Take-All' architecture that may underlie this type of processing \citep{DeCoensel, Koch1987}. Similarly, during attentive listening, certain sound events can be perceptually placed in the foreground and dominate most or all of the listener's focused attention, making other surrounding sounds effectively inaudible. This `inattentional deafness' takes place when (human) listeners attend to specific elements of a sound scene, but they can be completely oblivious to other sounds that are otherwise clearly audible \citep{Koreimann, Dalton}.

\subsection{Cognitive and other high-level mechanisms}

Information loss may occur despite successful auditory stream analysis and focused attention. This can happen because the sounds cannot be categorised by listeners (see \citealp{Pylyshyn2001}), the sounds are not represented in long-term memory in any accessible form \citep{McAdams1993}, or the sound event context is unclear and not helpful for determining their identity with certainty \citep{Ballas1993}. Unfamiliarity also seems to require more resources, as coding of unknown sounds may be less efficient compared to known sounds \citep{Attias,Fernald}. 

Sound processing and perception depend on different forms of auditory memory. Such are the real-time generation of streams from sequences of sounds (short-term auditory memory), promotion of sound identification (long-term auditory memory), or even the facilitation of forward masking and temporal integration \citep{Demany}. Additionally, basic computational processes such as comparison, discrimination, and adaptation also require the function of memory. Unsurprisingly, there are multiple possibilities for information loss because of memory-related phenomena, or rather, forgetting. This can be part of the normal operation of the auditory system \citep{McKeown}, or due to interference of one stimulus with another similar, ongoing one, so that the information about the former is lost \citep{Deutsch1970,Starr}. Informational bottlenecks can stem from limited working memory, or other executive processing measures, where processing of information over longer-time frames is required (e.g. \citealp{Baddeley, Cowan2010}).

If exacerbated, low motivation \citep{Balcetis}, physical or mental fatigue, or competing sensory information (e.g., \citealp{Stein, Lavie2010, Parmentier}), can affect performance and lead to switching of the listener's attention away from the auditory stream, in favour of an altogether different sensory input, or mind-wandering by task-irrelevant thoughts \citep{Lavie2010}. Similarly, listeners may physically react to what they hear (in decision-making) or engage in another behaviour that competes for attentional resources, or reorients perception, to optimise their situation through action \citep{Friston}. 

\section{Auditory information rate examples}
\label{speech}
\begin{figure}
		\centering
		\includegraphics[width=0.8\linewidth]{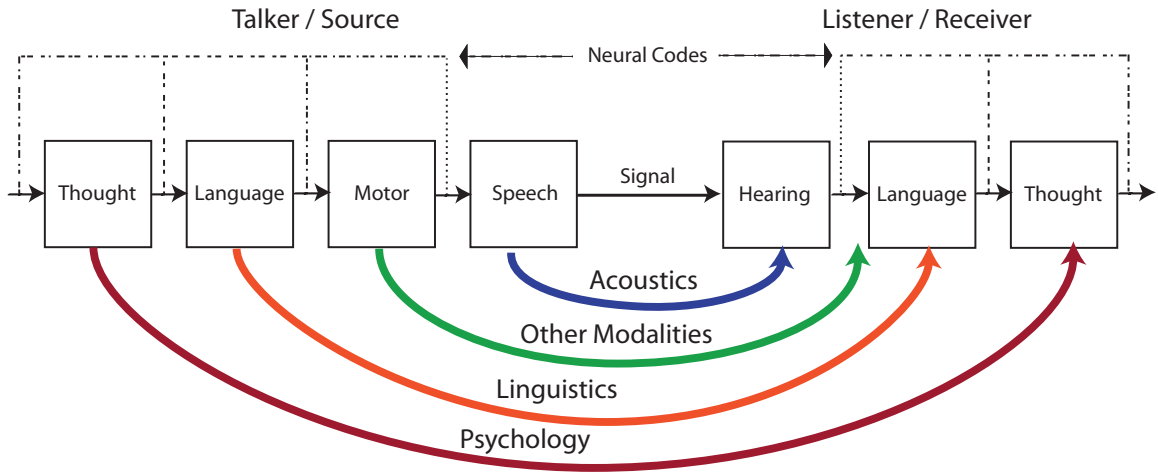}	
		\caption{Different levels of information flow of human speech. The lowest level (innermost in the figure) is the exact acoustic signal that the source transmits to the receiver (the talker says to the listener). This signal contains the highest information rate. One level up (outward), the motor system is responsible for modulating the airflow that results in speech production, which is maybe mirrored in speech perception as well \citep{Liberman, Poeppel}, among other parallel non-acoustic channels \citep[e.g.,][]{Sumby, Mcgurk}. The linguistic level is one level higher, which already has much lower and more efficient informational content than the acoustic signal \citep{Chomsky}. Finally, on an abstract psychological level, there is thought (goals, emotion, motivation, etc.) that generates / parses the messages to / from the linguistic modality. All information transfers that occurs within the brain are coded in intermediate neural signals, which may also include feedback and feedforward loops. Noise sources for all channels are not shown.}
		\label{fig:infoflow}
\end{figure}

Because of the multitude of auditory subsystems and special classes of sound, it is possible to tap the information flow at different places along the auditory pathways, or pick only a subset of acoustic situations and model them using a code that explains most of the effect. An example of the multiple processing levels, codes, and rates of information involved is given in Figure \ref{fig:infoflow}, which presents a speech-centric adaptation of Shannon's original communication system \citep{Shannon}. In the figure, the information communicated -- human language in this case -- may be analysed using codes in different levels of abstraction: acoustic, sensory-motor (or visual), neuronal, linguistic, and psychological. Some of these codes are much more readily operationalised than others \citep[cf.][]{Wiener}. 

The lowest level of information input to the hearing system is through sound waves that impinge on the outer ears. Each ear receives an approximately one-dimensional pressure wave reaching its eardrum. The acoustic signal can be sampled at more than twice its bandwidth, so typically at 44.1 kHz and bit depth of 16 bits/samples, for humans, the continuous external acoustic world is brought to a finite level of 705,600 bits/s per ear. A 4-11 times reduction of this rate is achieved with negligible loss of fidelity (depending on the sampled signal) by eliminating parts of sound, which would anyway be inaudible due to masking, among other methods of data compression \citep{Musmann}. However, the acoustic fidelity can be sacrificed to reduce the rate further, with negligible loss in speech intelligibility. In modern telephony standards, the signal bit rate can be compressed at 100-200 times the size of the uncompressed signal (4.75-23.85 kbits/s) \citep{Gibson}.  

Continuing to focus on speech communication as a particularly well-studied case -- if the acoustic packaging of language production is completely abstracted, then the linguistic message is much smaller than its full carrier signal \citep[p. 3-6]{Flanagan}. It has been empirically estimated using information processing response time task that the human cognition processes a maximum of 52 bits/s \citep{Martin}, and about 35 bits/s for speech using acoustic methods \citep{Leijon}. Even if these bounds are underestimated, they are still 3-4 orders of magnitude lower than the full acoustic signal, which contains much unnecessary information insofar as the high-level message is concerned, as well as useful information that may not be recoverable such as intonation, emotion, and voice-specific cues. The lossy aspect of language compression is most evident in written language, which is very low in bit rate compared to sound\footnote{Defining a relevant code for language is much more complicated than sound, though, since there are many speech elements that are not embodied in written language \citep[e.g.,][p. 1-10]{Beechey}.}. English text, for example, can be compressed by 50-75\% based on redundancy alone \citep{Shannon1951}, so that a speaking rate of three five-letter words per second can be textually transmitted at about 18 bit/s.

\section{Beyond hearing}
\label{OtherMods}
The unifying principle of the above auditory mechanisms -- being informationally lossy -- can be readily generalised to other modalities (cf., \citealp{Buracas}). The laboratory-based methods and relevance of the communication-theoretical framework can be translated to other senses, without loss of generality. Many of the mechanisms reviewed above, such as dynamic range, masking, masking release, scene analysis, selective attention, and memory, have parallels in other modalities. Additionally, natural sensory inputs are often multimodal, which can result in complex interactions, when cross-modal information is complementary, or in no interactions when they are completely independent \citep{Partan}. Whenever the modalities interact or are informationally redundant, some form of compression is inevitable. 

\section{Conclusion}
It was demonstrated mainly through human hearing -- that largely generalises to mammalian hearing -- how the brain handles the arbitrary sensory inputs of everyday environments, by being regularly engaged in lossy information processing. Cumulative and compressive information loss in hearing constitutes a unifying principle of many multilevel auditory effects, which have been usually studied in isolation in the hearing research literature. It is maintained that acoustic stimuli of unknown complexity are turned to processable, low-capacity output that the brain can use to enable survival, inform world models, and allow decision making and action, by avoiding overloading the system. This approach may be seen as a step toward parsimony in accounting for otherwise disjointed auditory phenomena, as well as toward real-world understanding thereof. These ideas can be readily generalised to other sensory modalities, as many of the underlying mechanisms have parallels across senses.

\section*{Acknowledgments}
The author would like to thank J\"org Buchholz and Gitte Keidser for supervising the lead-up to this work, and for help with formulating the conceptual model of Figure \ref{fig:LossModel}. Special thanks to Timothy Beechey, Kelly Miles, and Liviu Sigler for their useful comments. This work was supported by the Oticon Foundation and Macquarie University via an iMQRES PhD scholarship.

\bibliography{information_loss}
%\printbibliography[heading=bibintoc]

\end{document}